\documentclass[11pt]{article}
\usepackage{color}
\usepackage{tikz}
\usepackage{amssymb, amsmath}
\usepackage{epsfig}
\usepackage{graphicx}

\newtheorem{theorem}{Theorem}

\newcommand{\proof}{\noindent\textbf{Proof.~}}
\newcommand{\vep}{\varepsilon}
\newcommand{\qed}{\space\hfill\hspace*{\fill} $\vbox{\hrule\hbox{\vrule
height1.3ex\hskip1.3ex\vrule}\hrule}$\hss\vskip\topsep\relax}

\begin{document}

\title{An elementary approach to the option pricing problem}

\author{Nikolaos Halidias \\
{\small\textsl{Department of Mathematics }}\\
{\small\textsl{University of the Aegean }}\\
{\small\textsl{Karlovassi  83200  Samos, Greece} }\\
{\small\textsl{email: nikoshalidias@hotmail.com}}}

\maketitle

\begin{abstract}Our goal here is to discuss  the pricing problem of European and American options in
discrete time using elementary calculus so as to be an easy
reference for first year undergraduate students. Using the
binomial model we compute the fair price of European and American
options. We explain the notion of Arbitrage and the notion of the
fair price of an option using common sense. We give a criterion
that the holder can use to decide when it is appropriate to
exercise the option. We prove the put-call parity formulas for
both European and American options and we discuss the relation
between American and European options. We give also the bounds for
European and American options. We also discuss the portfolio's
optimization problem and the fair value in the case where the
holder can not produce the opposite portfolio.
\end{abstract}



\section{Introduction}
Our starting point was the paper \cite{Cox} in which the authors
introduce the binomial model and explain how one can use it to
evaluate the fair price of a European option. Our goal here is to
study the option pricing problem in discrete time using the
binomial method and basic calculus so as to be an easy reference
for first year undergraduate students.

There are many books that discuss the binomial model in a more
advanced setting, see for example \cite{Goodman}, \cite{Hull},
\cite{Kwok}, \cite{Musiela}, \cite{Pascucci1}, \cite{Pascucci2},
\cite{Shreve}, \cite{Williams}. Our aim here is to explain the
binomial method using elementary calculus but not losing any of
the mathematical accuracy.

We begin our discussion from the beginning, i.e. we describe
firstly how one can model the movement of an asset. Then, we
describe how can someone construct portfolios with prescribed
final and intermediate values and  we discuss both the European
and American type options. We also discuss the notion of the
Arbitrage and prove that the binomial model does not admit
Arbitrage under some suitable condition. We prove the put-call
parity formulas for both European and American options and we
discuss the relation between American and European options. We
give also the bounds for European and American options. We also
discuss the portfolio's optimization problem and the fair value in
the case where the holder can not produce the opposite portfolio.

 Suppose that our market
consists of one risky asset, say $S$, and one non-risky, say $B$,
with daily interest rate $r$. We consider for simplicity that we
have only one period, time zero and time one.

In time zero no one knows the value of the risky asset in time
one, i.e. no one knows $S_1$. How can we model this? We can for
example, study  the way that the asset behaved   the last, say,
one month and denote the average  of the percentage of got  up as
$u$ and the average of percentage of got down as $d$. Then we can
suppose that the risky asset will follow the same path in the
future and thus we can write schematically
\\[0.3cm]
\tikzstyle{bag} = [text width=2em, text centered] \tikzstyle{end}
= []
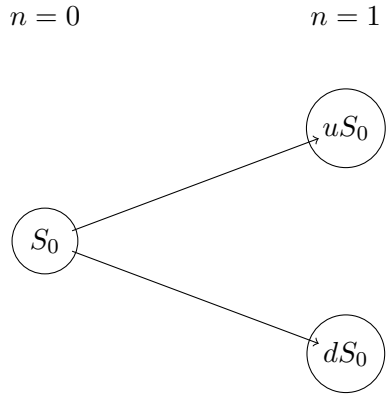
\begin{tikzpicture}[sloped]
\node [circle,draw] {$S_0$};
 \node [circle,draw] at (4,1.5) {$uS_0$};
 \node[circle,draw] at (4,-1.5) {$dS_0$} ;

  \node (a) at ( 0,0) [bag] {} ;
  \node (b) at ( 4,-1.5) [bag] {};
  \node (c) at ( 4,1.5) [bag] {};
  \node (d) at ( 8,-3) [bag] {};
  \node (e1) at ( 8,1) [bag] {};
  \node (e2) at ( 8,-1) [bag] {};
  \node (f) at ( 8,3) [bag] {};

  \node at (0,3) {$n=0$};
  \node at (4,3) {$n=1$};
  \draw [->] (a) to node [below] {} (b);
  \draw [->] (a) to node [above] {} (c);

\end{tikzpicture}
\section{Constructing a portfolio with prescribed final values}
 At time zero, someone can buy $a$ shares of the
risky asset and put the amount $b$ in the bank therefore
constructing a portfolio with initial value
\begin{eqnarray*}
V_0 = a S_0 + b
\end{eqnarray*}
If our time period is one day then after one day the value of the
portfolio will be
\begin{eqnarray*}
V_1^{u} = a(uS_0) + b(1+r)
\end{eqnarray*}
if the value of the asset will go up, and
\begin{eqnarray*}
V_1^{d} = a (dS_0) + b(1+r)
\end{eqnarray*}
if the value of the asset will go down. We can write it
schematically
\\[0.3cm]
\tikzstyle{bag} = [text width=2em, text centered] \tikzstyle{end}
= []
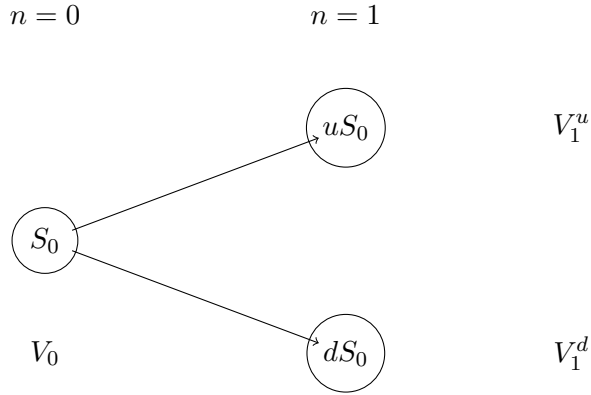
\begin{tikzpicture}[sloped]
\node [circle,draw] {$S_0$};
 \node [circle,draw] at (4,1.5) {$uS_0$};
 \node[circle,draw] at (4,-1.5) {$dS_0$} ;

  \node (a) at ( 0,0) [bag] {} ;
  \node (b) at ( 4,-1.5) [bag] {};
  \node (c) at ( 4,1.5) [bag] {};
  \node (d) at ( 8,-3) [bag] {};
  \node (e1) at ( 8,1) [bag] {};
  \node (e2) at ( 8,-1) [bag] {};
  \node (f) at ( 8,3) [bag] {};

  \node at (0,3) {$n=0$};
  \node at (4,3) {$n=1$};
  \node at (0,-1.5) {$V_0$};
  \node at (7,1.5) {$V_1^{u}$};
  \node at (7,-1.5) {$V_1^{d}$};
  \draw [->] (a) to node [below] {} (b);
  \draw [->] (a) to node [above] {} (c);

\end{tikzpicture}

Suppose now that we are given specific numbers $A,B$ and we are
asked to construct a portfolio $(a,b)$ such that, under the above
hypotheses, will have final values $V_1^{u} = A$ and $V_1^{d} =
B$. How much money $b$ we will have to put in the bank at time
zero and how many shares of the risky asset should we buy at time
zero? Schematically we have the following
\\[0.3cm]
\tikzstyle{bag} = [text width=2em, text centered] \tikzstyle{end}
= []
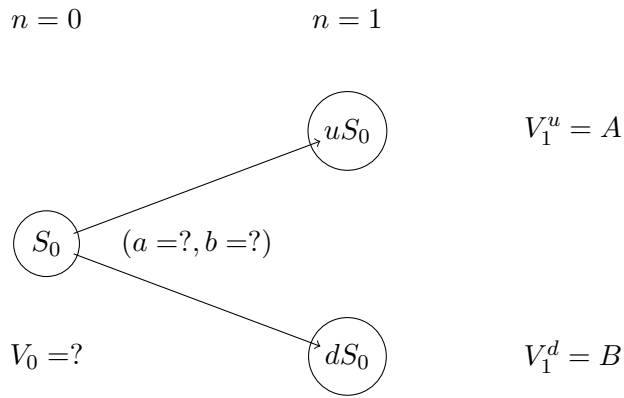
\begin{tikzpicture}[sloped]
\node [circle,draw] {$S_0$};
 \node [circle,draw] at (4,1.5) {$uS_0$};
 \node[circle,draw] at (4,-1.5) {$dS_0$} ;

  \node (a) at ( 0,0) [bag] {} ;
  \node (b) at ( 4,-1.5) [bag] {};
  \node (c) at ( 4,1.5) [bag] {};
  \node (d) at ( 8,-3) [bag] {};
  \node (e1) at ( 8,1) [bag] {};
  \node (e2) at ( 8,-1) [bag] {};
  \node (f) at ( 8,3) [bag] {};

  \node at (0,3) {$n=0$};
  \node at (4,3) {$n=1$};
  \node at (0,-1.5) {$V_0 = ?$};
  \node at (7,1.5) {$V_1^{u} = A$};
  \node at (7,-1.5) {$V_1^{d} = B$};
  \node at (2,0) {$(a=?,b=?)$};
  \draw [->] (a) to node [below] {} (b);
  \draw [->] (a) to node [above] {} (c);

\end{tikzpicture}

We have to solve two equations with two unknowns
\begin{eqnarray*}
a(uS_0) + b(1+r)  & = & A, \\
a(dS_0) + b(1+r) & = &  B
\end{eqnarray*}
The determinant of this system will be non zero if $S_0 \neq 0$,
$r \neq -1$ and $u > d$. Under the above hypotheses we have that
the solution of the system will be
\begin{eqnarray*}
a & = & \frac{A-B}{(u-d)S_0}, \\
b & = & \frac{Bu-Ad}{(u-d)(1+r)}
\end{eqnarray*}
Therefore the initial value of our portfolio has to be
\begin{eqnarray*}
V_0 = a S_0 +b = \frac{1}{1+r} \big(q A + (1-q)B \big)
\end{eqnarray*}
where $q = \frac{1+r-d}{u-d}$ and this results just if one
replaces $a,b$ from above to the \begin{eqnarray*} V_0 = aS_0+b
\end{eqnarray*} If $b\leq 0$ then that means that we have to borrow money
from the bank and if $a \leq 0$ means that we have to sell $a$
shares of the asset that we do not belong.

So, with the amount $V_0$ we have constructed a portfolio $(a,b)$
with final values $A,B$. Is there any chance to construct a
portfolio with final values $A+\vep_1$ and $B+\vep_2$ (with
$\vep_1
> $ and $\vep_2 > 0$) and initial value $V_0$? Let us find first the portfolio
$(a_1,b_1)$ with final values $A+\vep_1$ and $B+\vep_2$
\begin{eqnarray*}
a_1 & = & a + \frac{\vep_1-\vep_2}{(u-d)S_0}, \\
b_1 & = & b + \frac{\vep_2 u - \vep_1 d}{(u-d)(1+r)}
\end{eqnarray*}
Note that the portfolio $(a,b)$ have initial value $V_0$ and we
want also portfolio $(a_1,b_1)$ to have the same initial value but
bigger final values. Therefore, it must holds
\begin{eqnarray*}
\frac{\vep_1-\vep_2}{(u-d)} + \frac{\vep_2 u - \vep_1
d}{(u-d)(1+r)} = 0
\end{eqnarray*}
In other words, it must holds
\begin{eqnarray}
\vep_1(1+r-d) + \vep_2 (u-(1+r)) = 0
\end{eqnarray}

\section{Arbitrage and Smallest Initial Value}

Is there any portfolio $(a,b)$ with initial value $V_0 = 0$ and
final values
\begin{eqnarray*}
V_1^{u} > 0 \\
V_1^{d} \geq 0
\end{eqnarray*}
or
\begin{eqnarray*}
V_1^{u} \geq 0 \\
V_1^{d} > 0
\end{eqnarray*}
 If someone can construct such portfolios then he can borrow/put $b$ money
 from/to the bank to buy/sell  $a$ shares of the asset (again and again)
 and at the end he makes profit with zero initial capital and
 without any risk. Of course in real world there are not such
 portfolios so in our mathematical model we should exclude  such a
 situation which we call Arbitrage.

 \begin{theorem} The binomial model does not admit Arbitrage iff
 $0< d < 1+ r < u$
 \end{theorem}

 \proof
Let us suppose that $ 0 < d < 1+r < u$ holds. We construct a
portfolio $(a,b)$ such that
\begin{eqnarray*}
V_0 = aS_0 + b = 0
\end{eqnarray*}
so that $b = -a S_0$. Suppose now that
\begin{eqnarray*}
V_1^{u} & = & a(uS_0) + b(1+r) > 0 \\
V_1^{d} & = &a (dS_0) + b(1+r) \geq 0
\end{eqnarray*}
We will see now that in fact we have $a > 0$. We can write
\begin{eqnarray*}
V_1^{u} = auS_0 - aS_0(1+r) > 0
\end{eqnarray*}
Therefore we arrive at
\begin{eqnarray*}
aS_0(u-(1+r)) > 0
\end{eqnarray*}
Using the fact that $u > 1+r$ we obtain that $a > 0$. Substituting
the equality $b = - a S_0$ in the inequality $V_1^{d} \geq 0$ we
conclude that $d \geq 1+r$ but this is a contradiction. Using the
same arguments one concludes that it can not happen (if $d < 1+r <
u$)
\begin{eqnarray*}
V_1^{u}  \geq 0 \\
V_1^{d} > 0
\end{eqnarray*}

Conversely, suppose that the binomial model do not admit
Arbitrage. Consider all the possible portfolios with initial value
\begin{eqnarray*}
V_0  = aS_0 + b = 0
\end{eqnarray*}
If $V_1^{u} > 0$ then $V_1^{d} < 0$ otherwise $(a,b)$ is  an
Arbitrage opportunity.

By summing the inequalities $V_1^{u} > 0$ and $-V_1^{d} \geq 0$ we
get that $a > 0$. Using these inequalities and that $a > 0$ we get
the desired inequality, i.e. $d < 1+r < u$.

If $V_1^{u} < 0$ then $V_1^{d} >0$  otherwise $(-a,-b)$ is an
Arbitrage opportunity. The same arguments drives us to the same
conclusion.

If $V_1^{u} = 0$ then also $V_1^{d} = 0$ otherwise $(a,b)$ or
$(-a,-b)$ is an Arbitrage opportunity. By these two equalities we
conclude that $d = 1+ r = u$ and that mean that asset's value
remain  constant. \qed

From now on we will suppose that $0 < d < 1+r < u$ in order to
avoid Arbitrage in our model.

We have shown that for any $A,B$ one can construct a portfolio
$(a,b)$ with final values $V_1^{u} = A$ and $V_1^{d} = B$ under
the hypotheses that $u > d$, $r \neq -1$ and $S_0 \neq 0$. This is
called completeness of the model. What about the smallest initial
value of the portfolio with final values $A,B$? We have shown that
if someone want to construct a portfolio with final values
$A+\vep_1$ and $B+\vep_2$ then $\vep_1,\vep_2$ should satisfy
equation (1). Assuming that our model do not admit Arbitrage, then
equation (1) holds iff $\vep_1 = \vep_2 = 0$. Therefore, $V_0$ is
the smallest initial value for our portfolio with final values
$A,B$ if our model do not admit Arbitrage.

\section{Two period binomial model}
We can extend our results for a two period binomial model, i.e.
\\[0.3cm]
{\small \tikzstyle{bag} = [text width=2em, text centered]
\tikzstyle{end} = []
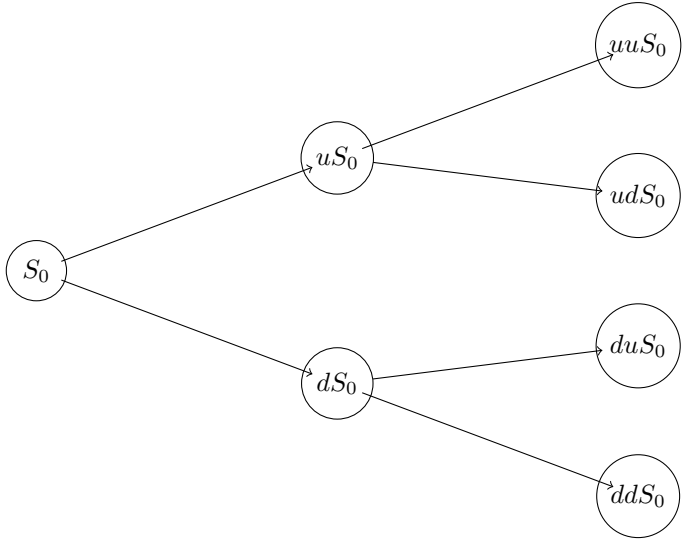
\begin{tikzpicture}[sloped]
\node [circle,draw] {$S_0$}; \node [circle,draw] at (4,1.5)
{$uS_0$}; \node[circle,draw] at (4,-1.5) {$dS_0$} ;
  \node [circle,draw] at ( 8,-3)  {$ddS_0$};
  \node [circle,draw] at ( 8,-1)  {$duS_0$};
  \node [circle,draw] at (8,1) {$udS_0$};
  \node [circle,draw] at ( 8,3)  {$uuS_0$};
  \node (a) at ( 0,0) [bag] {} ;
  \node (b) at ( 4,-1.5) [bag] {};
  \node (c) at ( 4,1.5) [bag] {};
  \node (d) at ( 8,-3) [bag] {};
  \node (e1) at ( 8,1) [bag] {};
  \node (e2) at ( 8,-1) [bag] {};
  \node (f) at ( 8,3) [bag] {};
  \draw [->] (a) to node [below] {} (b);
  \draw [->] (a) to node [above] {} (c);
  \draw [->] (c) to node [below] {} (f);
  \draw [->] (c) to node [above] {} (e1);
  \draw [->] (b) to node [below] {} (e2);
  \draw [->] (b) to node [above] {} (d);
\end{tikzpicture}}

If we do not have Arbitrage for the one period binomial model then
the same holds for the two period (and so on) binomial model. We
can also construct a portfolio and schematically have the
following
\\[0.3cm]
\tikzstyle{bag} = [text width=2em, text centered] \tikzstyle{end}
= []
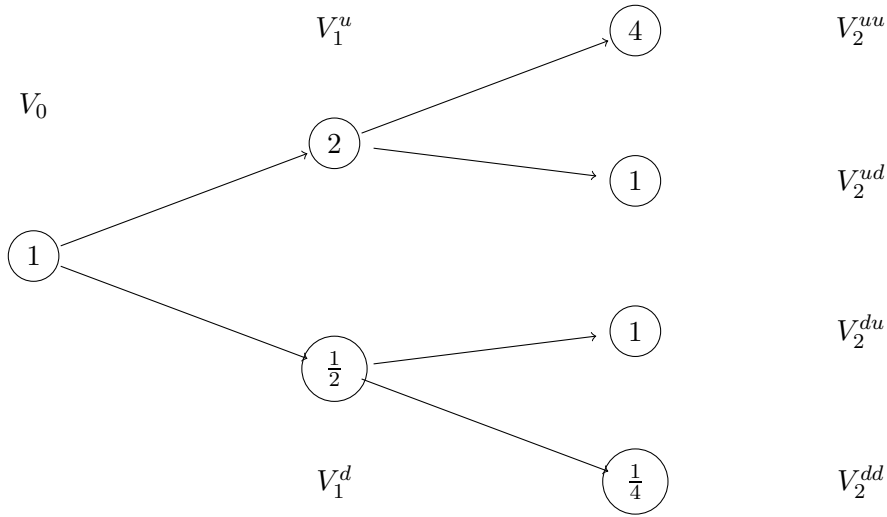
\begin{tikzpicture}[sloped]
\node [circle,draw] {1}; \node [circle,draw] at (4,1.5) {2};
\node[circle,draw] at (4,-1.5) {$\frac{1}{2}$} ;
  \node [circle,draw] at ( 8,-3)  {$\frac{1}{4}$};
  \node [circle,draw] at ( 8,-1)  {1};
  \node [circle,draw] at (8,1) {1};
  \node [circle,draw] at ( 8,3)  {4};
  \node (a) at ( 0,0) [bag] {} ;
  \node (b) at ( 4,-1.5) [bag] {};
  \node (c) at ( 4,1.5) [bag] {};
  \node (d) at ( 8,-3) [bag] {};
  \node (e1) at ( 8,1) [bag] {};
  \node (e2) at ( 8,-1) [bag] {};
  \node (f) at ( 8,3) [bag] {};
  \node at (11,3) {$V_2^{uu}$};
  \node at (11,1) {$V_2^{ud}$};
  \node at (11,-1) {$V_2^{du}$};
  \node at (11,-3) {$V_2^{dd}$};
  \node at (0,2) {$V_0$};
  \node at (4,3) {$V_1^{u}$};
  \node at (4,-3) {$V_1^{d}$};
  \draw [->] (a) to node [below] {} (b);
  \draw [->] (a) to node [above] {} (c);
  \draw [->] (c) to node [below] {} (f);
  \draw [->] (c) to node [above] {} (e1);
  \draw [->] (b) to node [below] {} (e2);
  \draw [->] (b) to node [above] {} (d);
\end{tikzpicture}
\\[0.2cm]
with
\begin{eqnarray*}
V_2^{uu} & = & a(uuS_0) + b(1+r)^2, \\
V_2^{ud} & = & a(udS_0) + b(1+r)^2, \\
V_2^{du} & = & a(duS_0) + b(1+r)^2, \\
V_2^{dd} & = & a(ddS_0) + b(1+r)^2
\end{eqnarray*}

Suppose now that we are given specified numbers
$A_2^{uu},A_2^{ud},A_2^{du},A_2^{dd}, A_1^{u},A_1^{d},A_0$ and we
are asked to construct the smallest  portfolio such that
\begin{eqnarray*}
V_2^{uu} & \geq & A_2^{uu}, \\
V_2^{ud} & \geq & A_2^{ud}, \\
V_2^{du} & \geq & A_2^{du}, \\
V_2^{dd} &  \geq &  A_2^{dd}, \\
V_1^{u} & \geq  & A_1^{u}, \\
V_1^{d} & \geq  & A_1^{d}, \\
V_0 & \geq & A_0
\end{eqnarray*}
Schematically we have
\\[0.3cm]
 \tikzstyle{bag} = [text width=2em, text centered]
\tikzstyle{end} = []
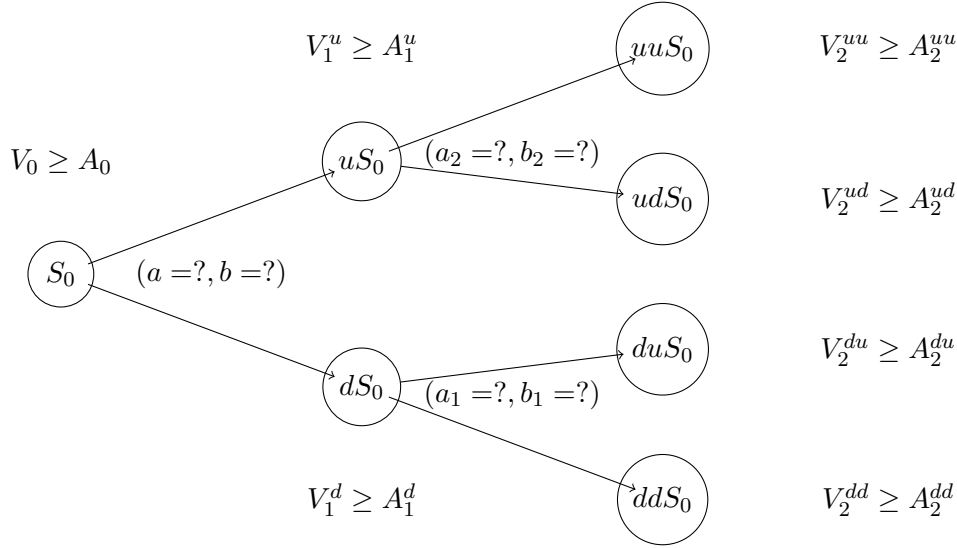
\begin{tikzpicture}[sloped]
\node [circle,draw] {$S_0$}; \node [circle,draw] at (4,1.5)
{$uS_0$}; \node [circle,draw] at (4,-1.5) {$dS_0$} ;
  \node [circle,draw] at ( 8,-3)  {$ddS_0$};
  \node [circle,draw] at ( 8,-1)  {$duS_0$};
  \node [circle,draw] at (8,1) {$udS_0$};
  \node [circle,draw] at ( 8,3)  {$uuS_0$};
  \node (a) at ( 0,0) [bag] {} ;
  \node (b) at ( 4,-1.5) [bag] {};
  \node (c) at ( 4,1.5) [bag] {};
  \node (d) at ( 8,-3) [bag] {};
  \node (e1) at ( 8,1) [bag] {};
  \node (e2) at ( 8,-1) [bag] {};
  \node (f) at ( 8,3) [bag] {};
  \node at (11,3) {$V_2^{uu} \geq A^{uu}_2$};
  \node at (11,1) {$V_2^{ud} \geq A^{ud}_2$};
  \node at (11,-1) {$V_2^{du} \geq A^{du}_2$};
  \node at (11,-3) {$V_2^{dd} \geq A^{dd}_2$};
  \node at (4,3) { $V_1^u \geq A^{u}_1$};
  \node at (4,-3) { $V_1^d \geq A^{d}_1$};
  \node at (6,1.6) {$(a_2 = ?,b_2 = ?)$};
  \node at (6,-1.6) {$(a_1 = ?,b_1 = ?)$};
  \node at (0,1.5) {$V_0 \geq A_0$};
  \node at (2,0) {$(a=?,b=?)$};
  \draw [->] (a) to node [below] {} (b);
  \draw [->] (a) to node [above] {} (c);
  \draw [->] (c) to node [below] {} (f);
  \draw [->] (c) to node [above] {} (e1);
  \draw [->] (b) to node [below] {} (e2);
  \draw [->] (b) to node [above] {} (d);
\end{tikzpicture}
\\[0.2cm]
How can we construct such a portfolio? We want $V_1^{u} \geq
A_1^{u}$ and also be such that at time 2 has the values
$A_2^{uu},A_2^{ud}$. Choosing,
\begin{eqnarray*}
V_1^{u} = \max \{ A_1^{u}, \frac{1}{1+r} \big(q A_2^{uu} + (1-q)
A_2^{ud} \big) \}
\end{eqnarray*}
we obtain the desired result. The same holds for $V_1^{d}$, i.e.
\begin{eqnarray*}
V_1^{d} = \max \{ A_1^{d}, \frac{1}{1+r} \big(q A_2^{du} + (1-q)
A_2^{dd} \big) \}
\end{eqnarray*}
Then, we choose $V_0 =\max \{A_0, \frac{1}{1+r} \big(q V_1^{u} +
(1-q)V_1^{d} \big) \}$. Next, we construct portfolios $(a,b)$,
$(a_2,b_2)$ and $(a_1,b_1)$.

It is clear that the initial value $V_0$ is the smallest value to
construct a portfolio with the specified requirements.

\section{European and American options}
A contract called European call option gives its holder the right
but not the obligation to purchase from the writer a prescribed
asset $S$ for a prescribed price $K$ at a prescribed time $T$ in
the future.

So, at time $T$ the holder of the option has the profit (if any)
$(S_T-K)^+$ and that is the amount of money that the writer of the
option has to pay at the expiration of the contract. It is obvious
that this kind of contract should have an initial cost for the
holder. What is the fair value $V_0$ of this contract? The writer,
at time $T$, using the amount $V_0$ should construct the profit of
the holder $(S_T-K)^+$. Therefore, if we consider our model (one
period model for example) the problem is to find $V_0,a,b$ to
construct a portfolio with specified final values. Schematically
we have the following,
\\[0.3cm]
{\small \tikzstyle{bag} = [text width=2em, text centered]
\tikzstyle{end} = []
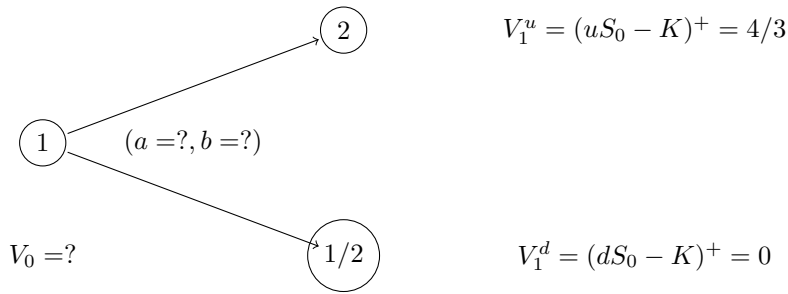
\begin{tikzpicture}[sloped]
\node [circle,draw] {$1$};
 \node [circle,draw] at (4,1.5) {$2$};
 \node[circle,draw] at (4,-1.5) {$1/2$} ;

  \node (a) at ( 0,0) [bag] {} ;
  \node (b) at ( 4,-1.5) [bag] {};
  \node (c) at ( 4,1.5) [bag] {};
  \node (d) at ( 8,-3) [bag] {};
  \node (e1) at ( 8,1) [bag] {};
  \node (e2) at ( 8,-1) [bag] {};
  \node (f) at ( 8,3) [bag] {};
\node at (8,1.5) {$V_1^u = (uS_0 - K)^+=4/3$}; \node at (8,-1.5)
{$V_1^d = (dS_0 - K)^+ = 0$}; \node at (2,0) {$(a=?,b=?)$}; \node
at (0,-1.5) {$V_0 = ?$};
  \draw [->] (a) to node [below] {} (b);
  \draw [->] (a) to node [above] {} (c);

\end{tikzpicture}}
\\[0.2cm]
with $S_0 = 1$, $u=2$ and $d=1/2$, $K = 2/3$ for example.
Therefore the problem can be solved as we have described before.
We have seen that if $d < 1+r < u$ then $V_0$, computed in this
way, is the smallest amount of money that the writer needs for
this contract in order to construct a portfolio that eliminates
the risk. Note that there is no path in which the writer loses or
earn money selling this contract. Moreover, the holder can lose
money but also can earn money from this contract. Any price above
$V_0$ will make sure profit (without risk) to the writer. What
about the case where the holder of the option do not exercise it
even in the case he has positive profit? Then this profit remain
to the writer. Is this an Arbitrage? In order to decide if it is
an Arbitrage or not we count only all the possible paths of the
asset. If for all possible paths of the asset the holder can
exercise in a way that the writer will have no sure profit then we
do not have Arbitrage. All these results can be extended to two
period models and in general to $n$ period models.

A contract called American call option gives its holder the right
but not the obligation to purchase from the writer a prescribed
asset for a prescribed price at any time until the expiration date
$T$. Therefore, the profit (if any) of the holder is $(S_t-K)^+$
where $t\leq T$ is the exercise time. The problem again is what is
the fair price $V_0$ of this contract. The writer should have
enough money for all the circumstances. Considering a two period
model the holder can exercise the option at times 0,1 as well at
time 2. Therefore, the writer should construct a portfolio which
has value at any time at least $(S_t-K)^+$ i.e. the holder's
profit (in order to eliminate the risk). So, the problem is to
specify the numbers $A_2^{uu},A_2^{ud},A_2^{du},A_2^{dd},
A_1^{u},A_1^{d},A_0$ and construct a portfolio such that

 \tikzstyle{bag} = [text width=2em, text centered]
\tikzstyle{end} = []
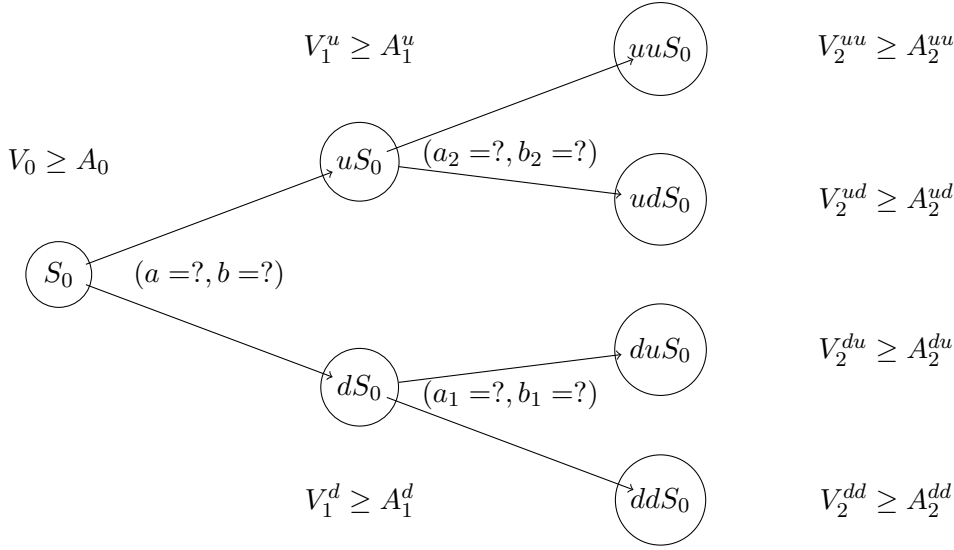
\begin{tikzpicture}[sloped]
\node [circle,draw] {$S_0$}; \node [circle,draw] at (4,1.5)
{$uS_0$}; \node [circle,draw] at (4,-1.5) {$dS_0$} ;
  \node [circle,draw] at ( 8,-3)  {$ddS_0$};
  \node [circle,draw] at ( 8,-1)  {$duS_0$};
  \node [circle,draw] at (8,1) {$udS_0$};
  \node [circle,draw] at ( 8,3)  {$uuS_0$};
  \node (a) at ( 0,0) [bag] {} ;
  \node (b) at ( 4,-1.5) [bag] {};
  \node (c) at ( 4,1.5) [bag] {};
  \node (d) at ( 8,-3) [bag] {};
  \node (e1) at ( 8,1) [bag] {};
  \node (e2) at ( 8,-1) [bag] {};
  \node (f) at ( 8,3) [bag] {};
  \node at (11,3) {$V_2^{uu} \geq A^{uu}_2$};
  \node at (11,1) {$V_2^{ud} \geq A^{ud}_2$};
  \node at (11,-1) {$V_2^{du} \geq A^{du}_2$};
  \node at (11,-3) {$V_2^{dd} \geq A^{dd}_2$};
  \node at (4,3) { $V_1^u \geq A^{u}_1$};
  \node at (4,-3) { $V_1^d \geq A^{d}_1$};
  \node at (6,1.6) {$(a_2 = ?,b_2 = ?)$};
  \node at (6,-1.6) {$(a_1 = ?,b_1 = ?)$};
  \node at (0,1.5) {$V_0 \geq A_0$};
  \node at (2,0) {$(a=?,b=?)$};
  \draw [->] (a) to node [below] {} (b);
  \draw [->] (a) to node [above] {} (c);
  \draw [->] (c) to node [below] {} (f);
  \draw [->] (c) to node [above] {} (e1);
  \draw [->] (b) to node [below] {} (e2);
  \draw [->] (b) to node [above] {} (d);
\end{tikzpicture}
\\[0.2cm]
Therefore we calculate $V_0$ as we have described and also
$(a,b)$, $(a_1,b_1)$ and $(a_2,b_2)$.  Using this amount of money
the writer will be sure that he will have enough money for all
possible paths of the asset (i.e. he can construct a portfolio
that eliminates the risk). There is no path that the writer will
make sure profit, because in each path the holder can exercise in
that a way that his profit equals portfolio's value.

Denoting by $X_n$ holder's profit at time $n$ when is the best
time to exercise the option? If the holder exercise at time $n$
then the profit will be in fact (considering the amount $V_0$ that
he had paid for this option) $X_n - V_0(1+r)^n$. Therefore, a
criterion is that the holder will exercise when $X_n >
V_0(1+r)^n$, i.e. in this case the holder will earn more money
than the case where he puts $V_0$ at time zero to a bank. If the
holder of the option choose to exercise it when $X_n  < V_n$ then
the writer has a profit. Is this an Arbitrage? No, it is not an
Arbitrage, because the notion of the Arbitrage is independent of
the choices of the holder. It depends only on all the possible
paths of the asset.

As an example consider an American call option with strike price
$K=5/2$, $N=3$, $u=2$, $d=1/2$ and $r=1/2$. Suppose that the asset
moves in the path $uu$ for the first two periods. The holder
should decide if he exercise at time $n=2$ the option or not.
 Making the
calculations the holder should choose to exercise at $n=2$ because
his profit is
$$(S_2^{uu} - K) - V_0(1+r)^2 = 3/2 - 0.48(1+1/2)^2 > 0$$ Note also that  the
value of the option at that time is $H_{2}^{uu} =V_2^{uu}= 2.44 >
X_{2}^{uu}=3/2$ and therefore the writer has a positive profit as
well. Of course the holder can choose to wait and if the asset go
up again he will make a larger profit but if the asset go down
then he loses all the money. If the holder of the option can sell
the option or he is able to sell a number of assets that he do not
own at time $n=2$ then he should decide (at this time) what is
preferable for him. In any case, at time $n=2$ he should take a
decision about his next move.

We should point out that if for some $n$ we have $ V_n^{A} = X_n $
and suppose further that the holder does not exercise at that time
then the writer has more money than he really needs to go to the
next step. Therefore, he can consume this extra amount of money
and invest the rest of them appropriately in order to eliminate
the risk. In this case let us denote the value of the portfolio as
$V_n^{AC}$. He also can put this amount of  money to the bank or
invest it on shares. In this case we denote the value of the
portfolio as $V_n^{A}$ and therefore it holds that $$V_n^{AC} \leq
V_n^{A}$$ Denote by $H_n$ the following sequence
\begin{eqnarray*}
H_n = \left\{%
\begin{array}{ll}
    X_N, & \mbox{ for } n=N,\\
    \max \{ X_n, \frac{1}{1+r} \left( q
H_{n+1}^{u}+(1-q)H_{n+1}^{d} \right) \}, & \mbox{ for } n=N-1,...,0\\
\end{array}%
\right.
\end{eqnarray*}
We say that $H_n$ is the fair value of the American option at time
$n$. Note that $V_n^{A} \geq H_n$.
\\[0.3cm]
{\tiny \tikzstyle{bag} = [text width=2em, text centered]
\tikzstyle{end} = []
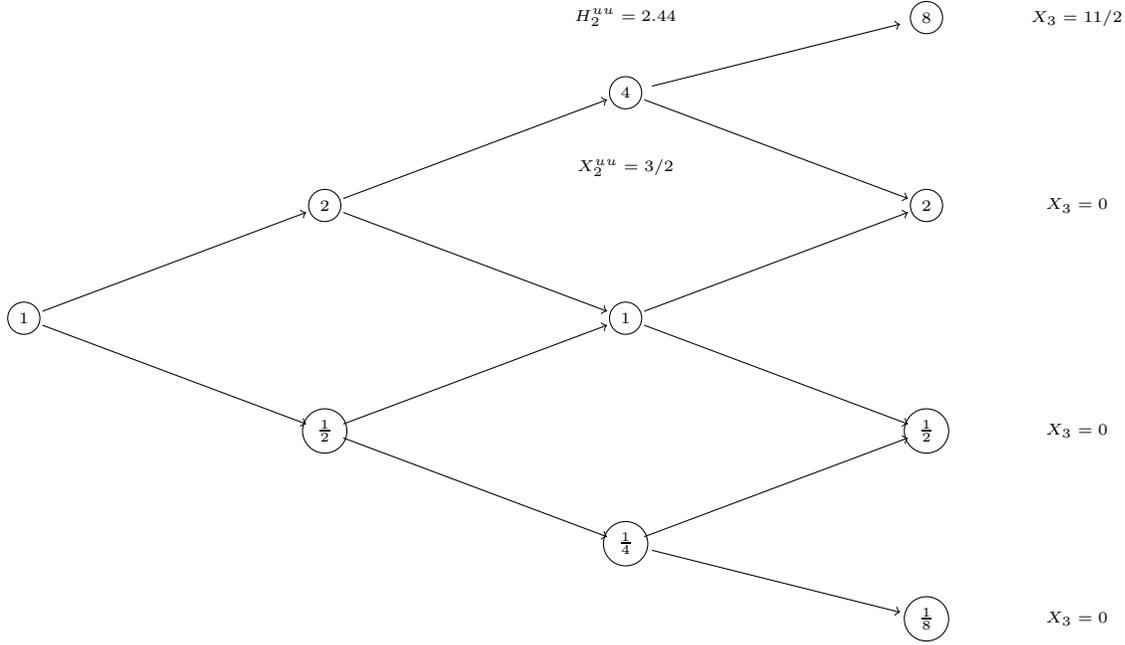
\begin{tikzpicture}[sloped]
\node [circle,draw] {1}; \node [circle,draw] at (4,1.5) {2}; \node
[circle,draw] at (4,-1.5) {$\frac{1}{2}$} ;
  \node [circle,draw] at ( 8,-3)  {$\frac{1}{4}$};
  \node [circle,draw] at ( 8,0)  {1};
  \node [circle,draw] at ( 8,3)  {4};
  \node [circle,draw] at (12,4) {8};
\node [circle,draw] at (12,1.5) {2};  \node [circle,draw] at
(12,-1.5) {$\frac{1}{2}$};\node [circle,draw] at (12,-4)
{$\frac{1}{8}$};
  \node (a) at ( 0,0) [bag] {} ;
  \node (b) at ( 4,-1.5) [bag] {};
  \node (c) at ( 4,1.5) [bag] {};
  \node (d) at ( 8,-3) [bag] {};
  \node (e) at ( 8,0) [bag] {};
  \node (f) at ( 8,3) [bag] {};
  \node (g) at ( 12,4) [bag] {};
  \node (h) at ( 12,1.5) [bag] {};
  \node (i) at ( 12,-1.5) [bag] {};
  \node (j) at ( 12,-4) [bag] {};

\node at (14,4){$X_3 = 11/2$}; \node at (14,1.5) {$X_3 = 0$};
\node at (14,-1.5) {$X_3=0$}; \node at (14,-4) {$X_3=0$}; \node at
(8,4) {$H_2^{uu} = 2.44$}; \node at (8,2) {$X_2^{uu} = 3/2$};
  \draw [->] (a) to node [below] {} (b);
  \draw [->] (a) to node [above] {} (c);
  \draw [->] (c) to node [below] {} (f);
  \draw [->] (c) to node [above] {} (e);
  \draw [->] (b) to node [below] {} (e);
  \draw [->] (b) to node [above] {} (d);
\draw [->] (f) to node [above] {} (g); \draw [->] (f) to node
[below] {} (h);\draw [->] (e) to node [above] {} (h);\draw [->]
(e) to node [below] {} (i);\draw [->] (d) to node [above] {}
(i);\draw [->] (d) to node [below] {} (j);
\end{tikzpicture}}
\\[0.3cm]

\section{Put-Call parity formulas, relations between European
and American options and bounds for options}

\subsection{European put - call parity} Consider a European call
option with strike price $K$ and the corresponding European put
option. Let us denote by $V_n^{E,call}$ and $V_n^{E,put}$ the
prices of the options working on an $N$-period binomial model. The
following formula holds,
\begin{eqnarray*}
V_n^{E,call}-V_n^{E,put} = S_n - K \frac{1}{(1+r)^{N-n}}, \quad
n=0,...,N
\end{eqnarray*}

Indeed, for $n=N$ we have
\begin{eqnarray*}
V_N^{E,call}-V_N^{E,put} = (S_N-K)^{+} - (K-S_N)^{+} = (S_N-K)^{+}
- (S_N-K)^{-} = S_N - K.
\end{eqnarray*}
Suppose that the formula holds for some $n$. We will show that it
holds also for $n-1$,
\begin{eqnarray*}
V_{n-1}^{E,call} - V_{n-1}^{E,put} & = & \frac{1}{1+r} \left(
q(V_n^{u,E,call} - V_n^{u,E,put}) +
(1-q)(V_n^{d,E,call}-V_n^{d,E,put}) \right) \\ & = & \frac{1}{1+r}
\left( q (uS_{n-1} - K \frac{1}{(1+r)^{N-n}}) + (1-q)(dS_{n-1} - K
\frac{1}{(1+r)^{N-n}}) \right) \\ & = & S_{n-1} - K
\frac{1}{(1+r)^{N-n+1}}
\end{eqnarray*}

\subsection{Relation between European and American options}

In general, it is easy to see that
\begin{eqnarray*}
V_n^{E} \leq H_n, \quad n=0,...N
\end{eqnarray*}
 Indeed, for
$n=N$ we have
\begin{eqnarray*}
V_N^{E}  = X_N = H_N
\end{eqnarray*}
where $X_N$ is the holder's profit at time $N$. Suppose that it
holds for some $n$, that is
\begin{eqnarray*}
V_n^{E} \leq H_n
\end{eqnarray*}
We will show that it holds also for $n-1$. We can write
\begin{eqnarray*}
H_{n-1} & = & \max \{ X_{n-1}, \frac{1}{1+r} \left( q H_n^{u} +
(1-q) H_n^{d} \right) \} \\ & \geq & \max \{ X_{n-1},
\frac{1}{1+r} \left( q V_n^{u,E} + (1-q) V_n^{d,E} \right) \} \\ &
= & \max \{ X_{n-1}, V_{n-1}^{E} \} \\ & \geq & V_{n-1}^{E}
\end{eqnarray*}

Consider now the case where we have a European option with strike
price $K$ and the corresponding American option in a $N$-period
model. If we speak about call options and $r \geq 0$ then we will
prove that
\begin{eqnarray*}
V_n^{E, call} = H_n^{call}  \quad \mbox{ (when } r \geq 0), \quad
n=0,...,N
\end{eqnarray*}
We need first to prove that
\begin{eqnarray*}
X_{n} \leq V_{n}^{E,call}, \quad n=0,...,N
\end{eqnarray*}
We will prove this by induction. For $n = N$ it is obvious, we
suppose that it holds for some $n$ and we will prove that it also
holds for $n-1$.
 To do so we  work as follows
 \begin{eqnarray*}
 S_{n-1} - K & = & \frac{(qu+(1-q)d)}{1+r}(S_{n-1} -
K) \\ &  = & \frac{1}{1+r}\left(q(uS_{n-1}-K)  +
(1-q)(dS_{n-1}-K)\right)\\& &  + \frac{1}{1+r}( qK(1-u)+(1-q)K(1-d))\\
& \leq
& \frac{1}{1+r}(q(uS_{n-1}-K)^{+} + (1-q)(dS_{n-1}-K)^{+}) \\ & & + \frac{1}{1+r}( qK(1-u)+(1-q)K(1-d))\\ & \leq & \frac{1}{1+r} (q V_n^{u,E,call}+ (1-q)V_n^{d,E,call}) \\
& = & V_{n-1}^{E,call}
\end{eqnarray*}
 where we have used the fact that $1+r = qu +(1-q)d$ and that
\begin{eqnarray*}
qK(1-u)+(1-q)K(1-d) = K-K(1+r) \leq 0
\end{eqnarray*}
Because $V_{n-1}^{E,call} \geq 0$ we have also that
\begin{eqnarray*}
X_{n-1} \leq V_{n-1}^{E,call}, \quad n=1,...,N
\end{eqnarray*}

Now we are ready to prove that $V_n^{E,call} = H_n^{call}$ by
induction. For $n = N$ is obvious so we suppose that it holds for
some $n$ and we will prove that it holds also for $n-1$.

Indeed,
\begin{eqnarray*}
H_{n-1}^{call} & = & \max \{ X_{n-1}, \frac{1}{1+r} \left( q
H_n^{u,call} + (1-q)H_n^{d,call} \right) \}\\ & = & \max \{
X_{n-1}, \frac{1}{1+r} \left( q V_n^{u,E,call} +
(1-q)V_n^{d,E,call} \right) \} \\ & = & \max \{ X_{n-1},
V_{n-1}^{E,call} \} \\ & = & V_{n-1}^{E,call} \end{eqnarray*}
Therefore, at any time $n$ there are no extra money for the writer
to consume and thus $$V_n^{A,call} = V_n^{AC,call}$$

Furthermore, if we speak about put options and $r=0$ then we also
have that
\begin{eqnarray*}
V_n^{E,put} = H_n^{put} \quad \mbox{ (when } r=0), \quad n=0,...,N
\end{eqnarray*}
We will first prove that $V_n^{E,put} \geq  X_n$ for $n=0,...,N$.
For $n=N$ it is obvious so we assume that it holds for some $n$
and we will prove it also for $n-1$. We work as follows
 \begin{eqnarray*}
K-S_{n-1} & = & K-dS_{n-1} + S_{n-1}(d-1) \\ & = &
(1-q)(K-dS_{n-1}) + q(K-dS_{n-1}) + S_{n-1}(d-1) \\ & = &
(1-q)(K-dS_{n-1}) +q(K-uS_{n-1})\\ & &   + qS_{n-1}(u-d) + S_{n-1} (d-1) \\
& \leq & (1-q)(K-dS_{n-1})^{+} + q(K-uS_{n-1})^{+} \\ & & +qS_{n-1}(u-d) +S_{n-1}(d-1) \\
& \leq  & V_{n-1}^{E,put}
\end{eqnarray*}
where we have used the fact that  $q = \frac{1-d}{u-d}$ so that
\begin{eqnarray*} S_{n-1} (d-1) = -qS_{n-1}(u-d)  \end{eqnarray*}
Note that, $V_{n-1}^{E,put} \geq 0$ therefore we also have
\begin{eqnarray*}
X_{n-1} \leq V_{n-1}^{E,put}, \quad n=1,...,N-1.
\end{eqnarray*}

Now, by induction it is easy to prove that $H_n^{put} =
V_n^{E,put}$. Indeed, for $n=N$ it is obvious and if it holds for
some $n$ then we will prove that it also holds for $n-1$.
Therefore
\begin{eqnarray*}
H_{n-1}^{put} & = & \max \{ X_{n-1}, \left( q H_n^{u,put} + (1-q)
H_n^{d,put} \right) \} \\ & = & \max \{ X_{n-1}, \left(q
V_n^{u,E,put} + (1-q) V_n^{d,E,put} \right) \} \\ & = & \max \{
X_{n-1}, V_{n-1}^{E,put} \} \\ & = &  V_{n-1}^{E,put}
\end{eqnarray*}
Finally, since there are no extra money at any time $n$ then
\begin{eqnarray*}
V_{n}^{AC,put} = V_n^{A,put}, \quad \mbox{ (when } r=0).
\end{eqnarray*}

\subsection{American put - call parity} Consider now an American
call option with strike price $K$ and the corresponding put
option. The following inequality holds,
 for a $N$-period binomial model, when $r \geq 0$,
\begin{eqnarray*}
 H_n^{call} - H_n^{put} \leq S_n - K
\frac{1}{(1+r)^{N-n}}, \; \; n=0,...,N
\end{eqnarray*}
To show this we note that $H_n^{ call} = V_n^{E,call}$ and
$V_n^{E,put} \leq H_n^{put}$ and  therefore
\begin{eqnarray*}
H_n^{call} - H_n^{put} \leq V_n^{E,call} - V_n^{E,put} = S_n - K
\frac{1}{(1+r)^{N-n}}
\end{eqnarray*}

The following inequality also holds,
\begin{eqnarray*}
S_n-K \leq H_n^{call} - H_n^{put}, \quad n=0,...,N
\end{eqnarray*}

We will show this inequality by induction. For $n=N$ we have
\begin{eqnarray*}
H_N^{call}  - X_N^{put} =  S_N-K
\end{eqnarray*}
Suppose that we have
\begin{eqnarray*}
H_n^{call} - H_n^{put} \geq S_n-K
\end{eqnarray*}
and we will show that
\begin{eqnarray*}
H_{n-1}^{call} - H_{n-1}^{put} \geq S_{n-1} - K
\end{eqnarray*}
Recall that
\begin{eqnarray*}
H_{n-1}^{put} & = & \max \{ (K-S_{n-1})^{+}, \frac{1}{1+r}
\left( q H_n^{u,put} + (1-q) H_n^{d,put} \right)  \} \\
H_{n-1}^{call} & = & \max \{ (S_{n-1}-K)^{+}, \frac{1}{1+r} \left(
q H_n^{u,call} + (1-q) H_n^{d,call} \right) \}
\end{eqnarray*}
Therefore
\begin{eqnarray*}
& &  H_{n-1}^{call} - H_{n-1}^{put} \\  & = & H_{n-1}^{call} + \min \Big\{ - (S_{n-1} -K)^{-}, \frac{-1}{1+r}\left( q H_n^{u,put} + (1-q) H_n^{d,put} \right)  \Big\} \\
& = & \min \Big\{ H_{n-1}^{call}- (S_{n-1}
-K)^{-},H_{n-1}^{call}-\frac{1}{1+r}\left( q H_n^{u,put} + (1-q) H_n^{d,put} \right)  \Big\} \\
 &\geq & \min \Big\{ (S_{n-1}-K), \frac{\left( q (
H_{n}^{u,call} -
H_n^{u,put}) + (1-q) ( H_{n}^{d,call} - H_n^{d,put}) \right)}{1+r}  \Big\} \\
& \geq & \min \Big\{ (S_{n-1}-K), \frac{\left( q(uS_{n-1}-K) +
(1-q)(dS_{n-1}-K) \right)}{1+r}  \Big\} \\ & \geq & S_{n-1}-K
\end{eqnarray*}
We have used the obvious inequalities
\begin{eqnarray*}
 H_{n-1}^{call} \geq (S_{n-1}-K)^{+} \mbox{ and } H_{n-1}^{call} \geq
\frac{1}{1+r} \left( q H_n^{u,call} + (1-q) H_n^{d,call} \right)
\end{eqnarray*}

To sum up we have proved the following inequalities,
\begin{eqnarray*}
S_n - K \leq  H_n^{call} - H_n^{put} \leq S_n - K
\frac{1}{(1+r)^{N-n}}, \;  \; n=0,...,N
\end{eqnarray*}

\subsection{Bounds for options}

We will show that
\begin{eqnarray*}
V_n^{E,call} = H_n^{call} \leq S_n
\end{eqnarray*}
For $n=N$ we have
\begin{eqnarray*}
V_N^{E,call} = (S_N-K)^{+} \leq S_N
\end{eqnarray*}
We suppose that
\begin{eqnarray*}
V_n^{E,call} \leq S_n
\end{eqnarray*}
and we will show that
\begin{eqnarray*}
V_{n-1}^{E,call} \leq S_{n-1}
\end{eqnarray*}
We have that
\begin{eqnarray*}
V_{n-1}^{E,call} & = & \frac{1}{1+r} \left( q V_{n}^{u,E,call} +
(1-q) V_n^{d,E,call} \right) \\ & \leq & \frac{1}{1+r} \left( q
uS_{n-1} + (1-q) dS_{n-1} \right) \\ & = & S_{n-1}
\end{eqnarray*}

Next we will show that
\begin{eqnarray*}
V_n^{E,call} \geq S_n - K\frac{1}{(1+r)^{N-n}}
\end{eqnarray*}
For $n=N$ we have that
\begin{eqnarray*}
V_N^{E,call} = (S_N-K)^{+} \geq S_N - K
\end{eqnarray*}
Suppose that we have
\begin{eqnarray*}
V_n^{E,call} \geq S_n-K\frac{1}{(1+r)^{N-n}}
\end{eqnarray*}
We will show that
\begin{eqnarray*}
V_{n-1}^{E,call} \geq S_{n-1}-K\frac{1}{(1+r)^{N-n+1}}
\end{eqnarray*}
We have that
\begin{eqnarray*}
V_{n-1}^{E,call} &  = & \frac{1}{1+r} \left( q V_n^{u,E,call} +
(1-q)V_{n}^{d,E,call} \right) \\ & \geq &  \frac{1}{1+r} \left( q
(uS_{n-1}-K\frac{1}{(1+r)^{N-n}}) + (1-q)(dS_{n-1} - K
\frac{1}{(1+r)^{N-n}}) \right) \\ & = &
S_{n-1}-K\frac{1}{(1+r)^{N-n+1}}
\end{eqnarray*}

Therefore we have proved so far that
\begin{eqnarray*}
S_{n}-K\frac{1}{(1+r)^{N-n}} \leq V_n^{E,call} = H_n^{call} \leq
S_n
\end{eqnarray*}

Next we will show that
\begin{eqnarray*}
K \frac{1}{(1+r)^{N-n}} - S_n \leq V_n^{E,put} \leq K
\frac{1}{(1+r)^{N-n}}
\end{eqnarray*}
For $n=N$ obviously we have that
\begin{eqnarray*}
K-S_N \leq V_N^{E,put} \leq K
\end{eqnarray*}
Suppose that it holds for some $n$, namely,
\begin{eqnarray*}
K \frac{1}{(1+r)^{N-n}} - S_n \leq V_n^{E,put} \leq K
\frac{1}{(1+r)^{N-n}}
\end{eqnarray*}
We will show that is also holds for $n-1$ that is
\begin{eqnarray*}
K \frac{1}{(1+r)^{N-n+1}} - S_{n-1} \leq V_{n-1}^{E,put} \leq K
\frac{1}{(1+r)^{N-n+1}}
\end{eqnarray*}
We have that
\begin{eqnarray*}
V_{n-1}^{E,put} & = &  \frac{1}{1+r} \left( q V_n^{u,E,put} +
(1-q)V_n^{d,E,put} \right)\\ & \leq & K \frac{1}{(1+r)^{N-n+1}}
\end{eqnarray*}
while
\begin{eqnarray*}
V_{n-1}^{E,put} & = &  \frac{1}{1+r} \left( q V_n^{u,E,put} +
(1-q)V_n^{d,E,put} \right) \\ & \geq & \frac{1}{1+r} \left( q (K
\frac{1}{(1+r)^{N-n}} - uS_{n-1}) + (1-q) (K \frac{1}{(1+r)^{N-n}}
- dS_{n-1}) \right) \\ & =  & K \frac{1}{(1+r)^{N-n+1}} - S_{n-1}
\end{eqnarray*}

Finally for American puts we have obviously that
\begin{eqnarray*}
(K-S_n)^{+} \leq H_n^{put}
\end{eqnarray*}
We will also show that
\begin{eqnarray*}
H_n^{put} \leq K
\end{eqnarray*}
For $n=N$ it is obvious. We suppose that it holds for some $n$ and
we will show that
\begin{eqnarray*}
H_{n-1}^{put} \leq K
\end{eqnarray*}
Indeed,
\begin{eqnarray*}
H_{n-1}^{put} & = & \max \{ (K-S_{n-1})^{+}, \frac{1}{1+r} \left(
q H_n^{u,put} + (1-q) H_n^{d,put} \right) \} \\ & \leq & \max \{
(K-S_{n-1})^{+}, \frac{1}{1+r} K \} \\ & \leq & K
\end{eqnarray*}

All the above relations and inequalities holds also in the
continuous case i.e.  as $n\to \infty$, in the spirit of
\cite{Jiang} (see also a more detailed  discussion of the
continuous case in \cite{Karatzas}). One can also prove, using
first year calculus, that the prices for European options
converges to the solution of the famous Black-Scholes-Merton
formula, see for example \cite{Pascucci2}, prop. 2.50.

We have proved all the above relations for the prices that the
binomial model produces, but using Arbitrage arguments one can
show that the same relations hold for the true prices, see for
example \cite{Hull}.

\section{Portfolio Optimization}

Suppose that we are given the amount of $V_0$ and we are able to
construct a portfolio putting money in the bank and buying a
number of assets. At time zero, our portfolio is
\begin{eqnarray}
V_0 = a S_0 + b
\end{eqnarray}
Suppose that we are working in an one period binomial model and
our problem is how much money we will put in the bank and how many
assets we should buy in order to maximize our portfolio at time 1.

Given a number $p \in (0,1)$  we want to maximize the quantity
\begin{eqnarray*}
pV_1^{u} + (1-p) V_1^{d}
\end{eqnarray*}
where
\begin{eqnarray*}
V_1^{u} & = & auS_0 + b(1+r), \\
V_1^{d} & = & adS_0 + b(1+r)
\end{eqnarray*}
Of course we want the following inequalities to hold
\begin{eqnarray*}
V_1^{u} & \geq & 0 \\
V_1^{d} & \geq & 0
\end{eqnarray*}
and using also (2) we arrive at the following constraint on $a$
\begin{eqnarray*}
-\frac{V_0(1+r)}{S_0(u-(1+r))}  \leq a \leq
\frac{V_0(1+r)}{S_0(1+r-d)}
\end{eqnarray*}
Note  that
\begin{eqnarray*}
pV_1^{u} + (1-p) V_1^{d} = aS_0\big(pu + (1-p)d - (1+r)\big) + V_0
(1+r)
\end{eqnarray*}
and therefore if $pu+qd > 1+r$ then we optimize our quantity if we
choose $a = \frac{V_0(1+r)}{S_0(1+r-d)}$ otherwise $a =
-\frac{V_0(1+r)}{S_0(u-(1+r))}$.

\section{Fair Value}
In section 5 we have estimated the smallest value of a European
option in order the writer to have enough money in any case. Is
that value a fair price? If the holder of the option can construct
the opposite portfolio (i.e. $(-a,-b)$) then indeed this is the
fair price because both the writer and the holder can construct
such a portfolios to eliminate the risk.

What about the case where the holder can not construct the
opposite portfolio? Then, only the writer has eliminate the risk
while the holder can lose or earn money buying this contract.
Intuitively speaking this value does not seem to be fair.

For a European option the expected profit of the holder is
\begin{eqnarray*}
\mathbb{E}^P (X) = pX_1^{u} + (1-p) X_1^{d}
\end{eqnarray*}
where $X_1^{u}$ is the profit if the asset goes up and $X_1^{d}$
is the profit if the asset goes down while $p$ is the probability
the asset goes up. Therefore a fair price could be the following
\begin{eqnarray*}
V_0 =\frac{1}{1+r} \min \Big\{ pX_1^{u}+(1-p)X_1^{d}, \;\;
qX_1^{u} + (1-q)X_1^{d} \Big\}
\end{eqnarray*}
where $q = \frac{1+r-d}{u-d}$.

If for a specific case $V_0 = \frac{1}{1+r} (qX_1^{u} +
(1-q)X_1^{d})$ then the writer can construct a portfolio to
eliminate the risk and if $V_0 = \frac{1}{1+r} (
pX_1^{u}+(1-p)X_1^{d})$ then the writer can put this amount of
money in the bank and so at time 1 this will be equal to the
holder's  expected profit or can construct a portfolio as we have
described  in Section 7.

For American type options a fair value could be
\begin{eqnarray*}
\min \{ V_0^q, \; V_0^p \}
\end{eqnarray*}
where $V_0^q$ is the amount of money that one needs to construct a
portfolio that eliminates the risk while $$V_0^p  = \max_{n}
\frac{1}{(1+r)^n} \mathbb{E}^P (X_n)$$ where $X_n$ is holder's
profit at time $n$.

\section{Conclusion} We described the European and American type options in
discrete time using basic calculus. We explain the notion of the
Arbitrage and we have seen that the fair price (when  the writer
and the holder can construct opposite portfolios) of an option is
in fact the smallest price that the constructed portfolio should
have as initial value in order the writer eliminate the risk and
this smallest price is closely connected with the no Arbitrage
criterion which is $d < 1+r < u$. Furthermore we have proposed a
criterion that the holder can have in mind to decide if he will
exercise the option at some specific time. We show the put-call
parity formulas for both European and American options and also
that the values of American and European call options coincides
when $r \geq 0$ while  the put options coincides  when $r=0$. We
also  give bounds for European and American call and put options.
 We have discussed the
portfolio's optimization problem and finally we discussed the
notion of the fair value of an option when the holder can not
construct an opposite portfolio to eliminate the risk.


\begin{thebibliography}{99}

\bibitem{Cox} {J. Cox - S. Ross - M. Rubinstein}, {\em Option
pricing: a simplified approach}, J. Financial Econ. 7, 229-264
(1979).



\bibitem{Goodman} {V. Goodman - J. Stamphli}, {\em The Mathematics
od Finance: Modelling and Hedging}, AMS, (2009).



\bibitem{Hull} {J. Hull}, {\em Options, Futures and Other
Derivatives}, Prentice Hall, (2008).


\bibitem{Jiang} {L. Jiang - M. Dai}, {\em Convergence of binomial
tree methods for European/American path-dependent options}, SIAM
J. Num. Anal, 42, pp. 1094-1109, (2004).





\bibitem{Karatzas} {I. Karatzas - S. Shreve}, { \em Methods of
Mathematical Finance}, Springer, (1998).


\bibitem{Kwok} {Yue-Kuen Kwok}, {\em Mathematical Models of
financial Derivatives}, Springer, (1998).


\bibitem{Musiela} {M. Musiela - M. Rutkowski}, { \em Martingale
Methods in Financial Modelling}, Springer, (2005).


\bibitem{Pascucci1} {A. Pascucci}, {\em PDE and Martingale Methods
in Option Pricing}, Springer, (2011).

\bibitem{Pascucci2} {A. Pascucci - W. Runggaldier}, {\em Financial
Mathematics, Theory and Problems for Multi-period Models},
Springer, (2012).



\bibitem{Shreve} {S. Shreve}, {\em Stochastic Calculus for Finance I: The Binomial Asset Pricing Model
}, Springer, (2004).



\bibitem{Williams} {R. J. Williams}, {\em Introduction to the
Mathematics of Finance}, AMS, (2006).

\end{thebibliography}
\end{document}